# Attention Based Molecule Generation via Hierarchical Variational Autoencoder

Divahar Sivanesan

**Abstract**

Molecule generation is a task made very difficult by the complex ways which we represent molecules computationally. A common technique used in molecular generative modeling is to use SMILES strings with recurrent neural networks built into variational autoencoders - but these suffer from a myriad of issues: vanishing gradients, long-range forgetting, and invalid molecules. In this work, we show that by combining recurrent neural networks with convolutional networks in a hierarchical manner, we are able to both extract autoregressive information from SMILES strings while maintaining signal and long-range dependencies. This allows for generations with very high validity rates on the order of 95% when reconstructing known molecules. We also observe an average Tanimoto similarity of .6 between test set and reconstructed molecules, which suggests our method is able to map between SMILES strings and their learned representations in a more effective way than prior works using similar methods.

# 1 Introduction

In recent years, Machine Learning has shown an unprecedented ability to learn complex relationships among abstract data given minimal prior knowledge. This has been particularly significant in the field of Natural Language Processing, where machine learning techniques are able to decompose entire paragraphs into analytically-useful numerical representations that can then be used to for downstream tasks such as sentiment analysis and question answering [1]. This facilitates a myriad of possible applications in fields that previously required tailored and fine-tuned solutions based on a significant amount of domain knowledge, for example, computational drug discovery [2].

Crucial to the utility of machine learning algorithms applied to chemical analysis is their ability to produce fixed-dimension and information dense numerical representations of molecules, which can be used in downstream tasks including property prediction, bio-activity prediction, and even conditional molecule generation. Historically, this has been done with molecular fingerprints, which are long and sparse bit vectors computed without machine learning techniques. Some research has focused specifically on producing better fingerprints which can be learned, allowing for representations tailored to specific problems like drug-protein binding affinity prediction [3]. We maintain the belief that although it is important how one chooses to numerically represent molecules, equally important are the tasks which are facilitated by the representation, and improvements in performance over classical encoding such as Morgan fingerprints.

We treat molecules as variable-length sequences of SMILES strings, which allow us to draw inspiration from NLP techniques [4]. We aim to create a Variational Autoencoder-like scheme capable of compressing molecule strings into *physically-relevant* encoded representations (henceforth *latent vectors*), and subsequently produce molecules that have a high Tanimoto similarity with the input molecule. There are other works which make use of VAEs for generating unique molecules like [5] and [6]. These works focus more on implementing previously known architectures with few modifications to molecule generation. In this paper, we present a novel hierarchical VAE architecture which has classical recurrent elements of NLP generative model schemes, as well as classical signal processing techniques to deal with the sequence data. molecules. Our contributions are as follows.

- We propose a novel hierarchical VAE with Attention which deals with the problem of generating long sequences.

- We use a Convolutional Neural Network that encodes the temporal information of the encoded sequence into low dimensional latent space.

# 2 Methods

Earlier work with variational autoencoders for molecule design typically used one of two encoding schemes - convolutional or recurrent [5]. There are pros and cons to each of these paradigms, but in our method, we cultivate an architecture that makes use of both, capturing the benefits of each scheme while mitigating many of the costs.

## 2.1 Recurrent Encoding

Recurrent encoding requires a global aggregation function in order to reduce the variable length tensor of hidden states to a fixed-length vector suitable for transformations and reparameterization, but we find in our experiments that these aggregations result in a large amount of information loss. We also observe a longer time to converge. However, the autoregressive nature of recurrent encoding is of particular importance for this task, since many SMILES strings contain common molecular sub-structures which could easily be generated autoregressively. For this reason, the first step in our algorithm is to encode a SMILES string autoregressively with a GRU.

Given a SMILES string $S$, $S_t$ is the token at position $t$. Each token in $S$ is tokenized with a vocabulary of 131 learnable embeddings, and passed to the recurrent encoder $\phi_{(e)}$.

$$h_{(e)t} = \phi_{(e)}(S_t, h_{(e)t-1}) \tag{1}$$

Where $h_{(e)t}$ is the hidden state of the recurrent encoder at time $t$. By combining all these hidden states $h_t$, we get a new tensor $\mathbf{H}$. This step of the procedure can be seen in figure 1A.

$$\mathbf{H} = \{h_0, h_1, ..., h_t\} \tag{2}$$

## 2.2 Variational Encoding

Our experiments involving PCA, which were originally done to show information loss when using aggregation during recurrent encoding, also showed a significant amount of redundant information in the hidden state tensors. To attempt to reduce the complexity of the encoding scheme, we borrowed ideas from the signal processing domain. The output of the recurrent encoder can be treated as a one dimensional signal with many channels (number of hidden dimensions of the recurrent encoder, in this

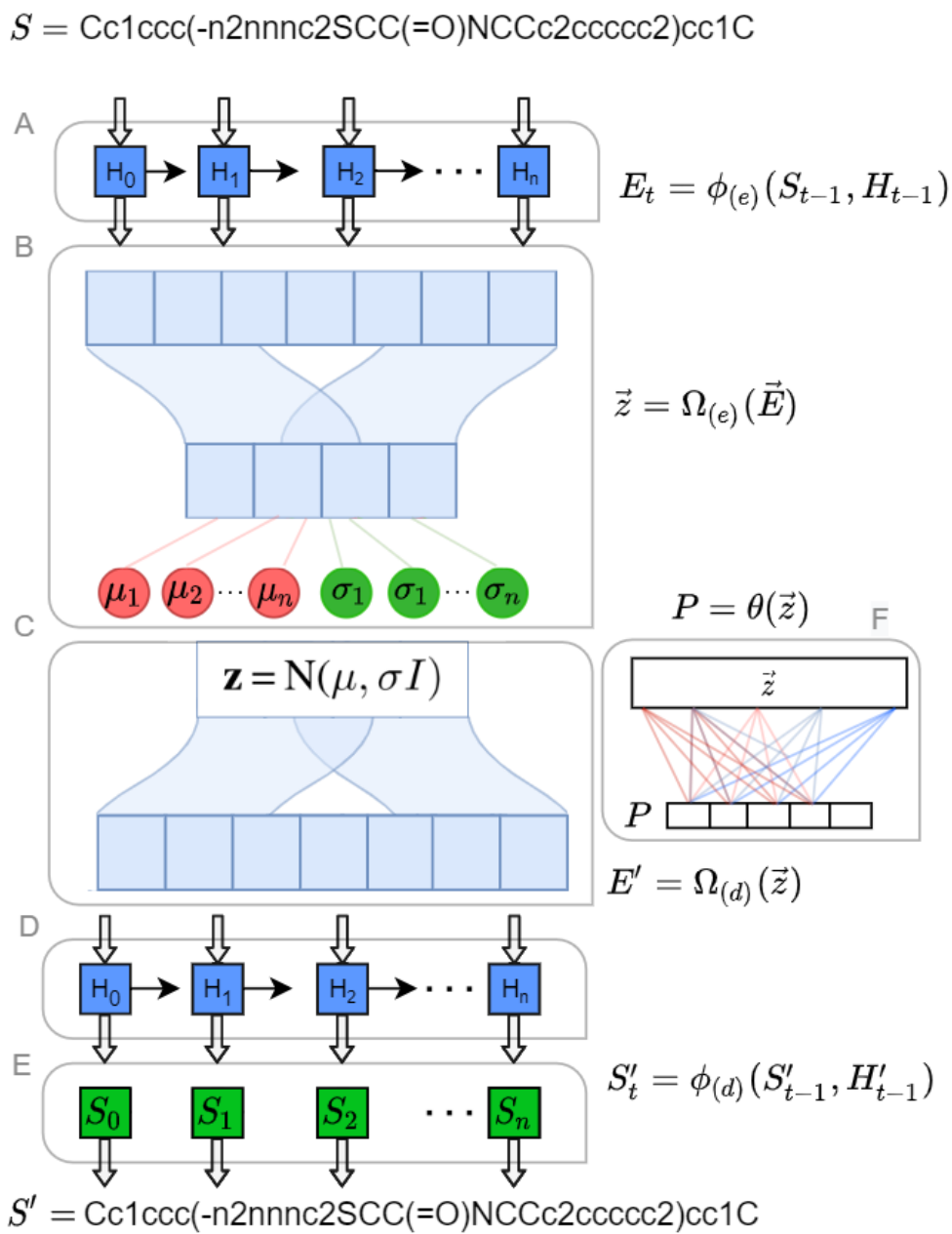

Figure 1. Full diagram of our model architecture pipeline

case).The purpose of the variational encoder is to summarize the signals in the hidden state tensor into a single information-dense vector.

The design of the variational encoder is fairly simple, consisting of only three convolutional layers with ReLU activations. After the third layer, the output is split down the middle, and the two sets of values are used to parameterize an isotropic Gaussian distribution as per the reparameterization trick. This makes up the second part of figure 1B, which notably lacks linear layers. Our experiments showed that the inclusion of a linear layer caused performance issues, which is consistent with our signal processing perspective that the latent vector should contain temporal information. We ran our experiments with a latent dimension of 128, 256, 512, and even 4096 in an attempt to overfit to our training set. We found best (most generalizable) results with a latent dimension of 256. We denote this second stage encoder $\Omega_{(e)}$, and it can be seen in figure 1B. Note that in the figure, the intermediate step between the distribution parameterization and convolutional layer output is *not* a linear layer.

### 2.3 Property prediction

The value in the latent representations comes, in large part, from the prediction of relevant properties. [8] quite recently reported that when training a transformer on 250 million proteins, a large transformer model was able to learn high-level information about proteins, despite having nothing but strings. They were able to show this by fine-tuning the model on downstream tasks, which converged quickly, demonstrating that the model mostly already knew how to model proteins effectively. We wish to learn similarly useful representations, but doing so without millions of training examples poses a challenge. By resorting to direct prediction of the properties that were deemed relevant for the downstream tasks, we are able to learn chemical significance with only 250,000 SMILES strings.

We do the property prediction the same way as Gomez. et. al. [5], except for a different set of properties. We take the latent representation $\vec{z}$ and pass it through two linear layers with ReLU activation functions, with 5 output nodes. We do min-max normalization for each property to a 0-1 scale, and observe that with or without a sigmoid activation, performance is usually about the same. We denote the property prediction head as $\theta(\vec{z})$, and choose to predict SA score, QED, TPSA, BertzCT, and molecular weight for each molecule. Gomez. et. al. [5] choose a much larger set of properties, which we reduce to these five for simplicity, particularly because QED is a quantity computed from many other molecular properties that are also relevant. We are able to analytically compute these properties with RDKit [7], which we then use as the ground truth. This part of the architecture is depicted in figure 1F.

### 2.4 Decoding

Our decoding scheme is similar to the inverse of the encoding scheme with a few modifications. In the first phase, we replace convolutional layers in the variational encoder with transposed convolutions, which upsample the latent vector by a factor of four. We call this portion of the decoder the *variational decoder*, and our experiments show that it is a large part of what is responsible for our performance improvements over current autoregressive molecular generative models. We denote it as $\Omega_{(d)}$.

$$E' = \Omega_{(d)}(z) \quad (3)$$

$z$ is the latent encoding of the sequence $E$, and $E'$ is the decoded sequence after passing through the network.

#### 2.4.1 Autoregressive Attention

In the second phase, we present an additional novel procedure. Our autoregressive decoding scheme differs from past works by incorporating a mechanism inspired by attention, which takes a tensor of all past embeddings $\vec{S}$ and a tensor of all past hidden states $\vec{H}$ from the GRU decoder, and does the following computation at each timestep;

$$Attention(E', \vec{H}, \vec{S}) = E' softmax(linear(\frac{\vec{H}\vec{S}^T}{\sqrt{\ell}})) \quad (4)$$

Where the operation between the variational decoder output, $E'$, and the attention weights is a hadamard product, and the operation between the hidden state and embedding tensors is a batch matrix multiplication with a transpose operation. The transpose operation seeks to remedy the issue of varying dimensions $\ell$ (corresponding to the current decoded sequence length) with matrix multiplication; the hidden state tensor of dimension $(N, \ell, hidden\ dimension)$ can be batch multiplied by the embedding tensor so long as we transpose the embedding tensor to be of shape $(N, \ell, embedding\ dimension)$. This allows us to not care how much of the sequence we have decoded so far - we will always have consistent shapes. We additionally include the factor of $\sqrt{\ell}$, as was done in Vaswani et. al. [8] to account for inflating values with longer sequences.

## 2.5 Training

### 2.5.1 Objective Function

The training of each module was carried out in parallel, so the loss function composed of three different terms.

- **Reconstruction Loss:** This made use of Cross Entropy function, with reduction as sum. There was no coefficient associated with this loss or another way to look at this is as having identity as coefficient.

- **Kullback Leibler Divergence:** This tried to regularize the latent space so that it's easy to generate new samples by sampling a new latent vector. The coefficient associated with this loss was $\beta$. Choosing the correct value for beta appeared to have a significant impact on eventual sample quality. We found a value of $1 \times 10^{-5}$ to be optimal.

- **Property Loss:** We optimized the property error by minimizing the Sum Squared Error. Another coefficient $\gamma$ was used to weigh this loss term. A value of $150$ gave the optimal results.

### 2.5.2 Teacher Forcing

No teacher forcing was used during training. We did experiment with using different ratios of teacher forcing, but none gave good results. One intuition on why teacher forcing is a bad idea while generating molecules is that with teacher forcing there is a high chance that the generated sequence will diverge during inference even if it makes a slightest mistake. The syntax we used for representing the molecules was SMILES, so even a slight mistake could render the whole sequence invalid syntactically. We did observe increased difficulty in converging when not using teacher forcing, but the difficulties tended to dissipate after about 2 epochs.

### 2.5.3 Optimization

We used the Adam optimizer for backpropagation. Betas used were $0.9$ and $0.999$ respectively, with a learning rate of $0.001$. To avoid the problem of exploding gradients, we used gradient clipping with a value of $0.5$. We found that a smaller batch size, in the range of 16-32, gave a faster convergence measured both by wall time and number of epochs.

### 2.5.4 Dataset

All our experiments were performed using the ZINC250k dataset [9] using 90% of the data, randomly sampled, as a training set, and 5% for the validation and test sets. Using RDKit [7], we computed the properties for each molecule in the database, we split the data, and saved the splits to separate files to ensure no overlap.

## 3 Results & Discussion

Our experiments aimed to evaluate the quality of the learned representations, as well as the ability of the model to go between a representation of a molecule and the molecule itself in both directions. This boils down to analysing the performance of three tasks;

- Molecule Recreation
- Molecule Generation
- Property Prediction

## 3.1 Molecule Reconstruction

We used two different metrics to measure the ability of the model to reconstruct a given sequence. The first of which is the sequence to sequence loss, which considers the cross entropy between two sequences. The other metric is the Tanimoto similarity, which is a well known chemical similarity score found (usually) by computing the Jaccard score for the Morgan Fingerprints of two molecules.

Of these two metrics, only categorical cross entropy is differentiable. We were forced to use it for optimization despite it being sub-optimal, due to the fact that SMILES strings are not unique with respect to molecules, meaning perfect Tanimoto scores can be achieved with imperfect accuracy. Another issue occurs when a sequence diverges from the original sequence during decoding, and then lags behind by one or more timesteps. It will render an extremely low accuracy, however there is a high probability that this original and reconstructed molecules will have high Tanimoto similarity.

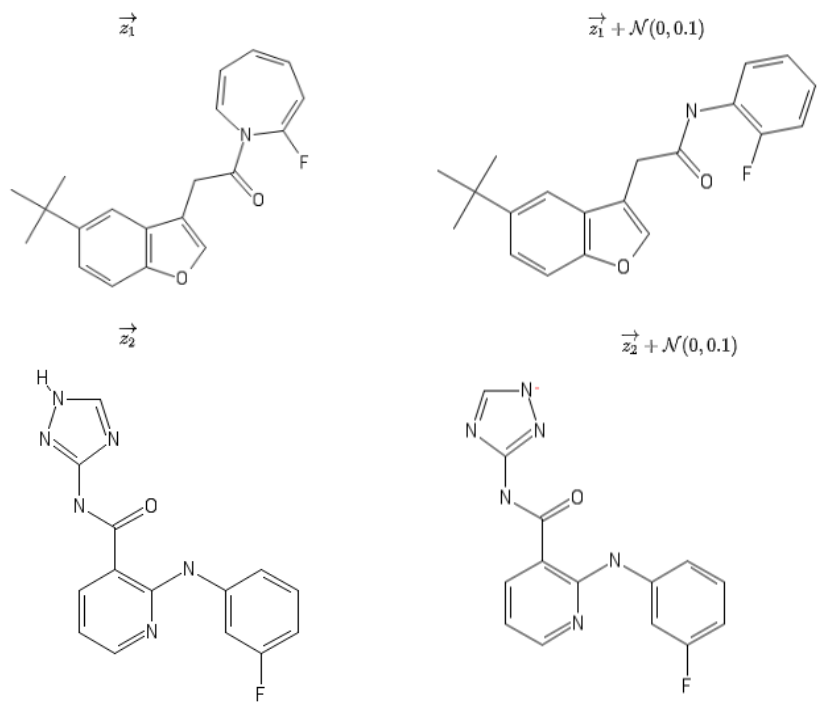

Figure 2. Training set molecules with latent vectors $\vec{z_1}$ and $\vec{z_2}$, their Gaussian-noise-added counterparts.

In the end, we found that our test set reconstructions averaged a Tanimoto similarity of .6, but this seemed to range from .55 to a little over .6 even with identical run configurations. As a further experiment, we added Gaussian noise to the latent tensor of our test set molecules, and computed the Tanimoto similarity between the decoded results and the original test set molecules. We found that with noise

## 3.2 Molecule Generation

When generating molecules unconditionally, we found that the largest issue was validity. There is no inherent rule that can be applied to our model that constrains it to produce only valid molecules; indeed, this has been a big problem in past works [5]. Our method was able to achieve a validity rate of close to 95% for reconstruction of known molecules, and a not insignificant 40% for unconditional sampling of the latent space.

We found that for generating new, unique, "good" molecules, best results are obtained when sampling around training set latent vectors by means of Gaussian noise. Not only are validity rates superior, but the property predictions are often more accurate, implying better representations than points far from training set observations. Additionally, we measured very high Tanimoto similarity, averaging around .80 between the training set molecules, and training set molecule latent vectors with noise. Depicted in Figure 2 are two examples.

### 3.2.1 Manifold of "good drugs"

We argue that within our latent space lies a manifold of "good drugs" which is learned involuntarily during training. Moving along this manifold would produce a very high rate of chemically valid molecules, but they would be similar to training set molecules as measured by Tanimoto similarity. We believe that with more training data, we can expand this manifold so that our latent space yields a higher proportion of valid molecules for random unconditional samples, as well as more chemically unique molecules. We maintain this belief after observing that the uniqueness of molecules sampled unconditionally tended to increase with dataset size.

## 3.3 Property Prediction

Although the numerical error is relevant in determining model performance, the original goal as stated in section 2.3 of property prediction is to attach physical significance to the latent space. We can encode our test set molecules into a large tensor of latent vectors, perform a dimensionality reduction technique to two dimensions, and assign the color to points based on scalar property values. To accomplish this, we try PCA and T-SNE and find similar results. We also try ground truth and predicted property values, and yet again find similar results. Figure 3 shows that our latent space is neatly organized according to these properties, meaning that property-based optimization is practically feasible.

## 4 Future Work

Our work appears to obtain a higher molecular validity rate than current autoregressive models based on SMILES strings, while also achieving a higher chemical similarity between source and reconstructed molecules. Our model also appears to learn a latent space with chemically significant characteristics, as we can see in Figure 3. This work can enable future work in targeted, conditional molecule design, be it for industrial or medical applications. Additionally, the architecture itself could be improved in a wide variety of ways. Although we are satisfied with our results, we believe the future of Machine Learning driven molecule design will be pioneered by incremental improvements on techniques like these.

## 5 Conclusion

We proposed a novel architecture for Sequence-to-Sequence VAEs, which makes use of the Attention mechanism to improve the generation quality of the molecules. Molecule reconstruction provided very promising results by both our target metrics, namely an average reconstruction Tanimoto similarity of .6 with a validity rate of over 95%. Other than reconstruction, the results on downstream tasks such as property prediction also produced results suggestive of a very physically meaningful latent space, which is critical for future work. This method is just a starting point and can be improved on many accounts. Furthermore, we believe applications to other domains of this architecture can also prove useful.

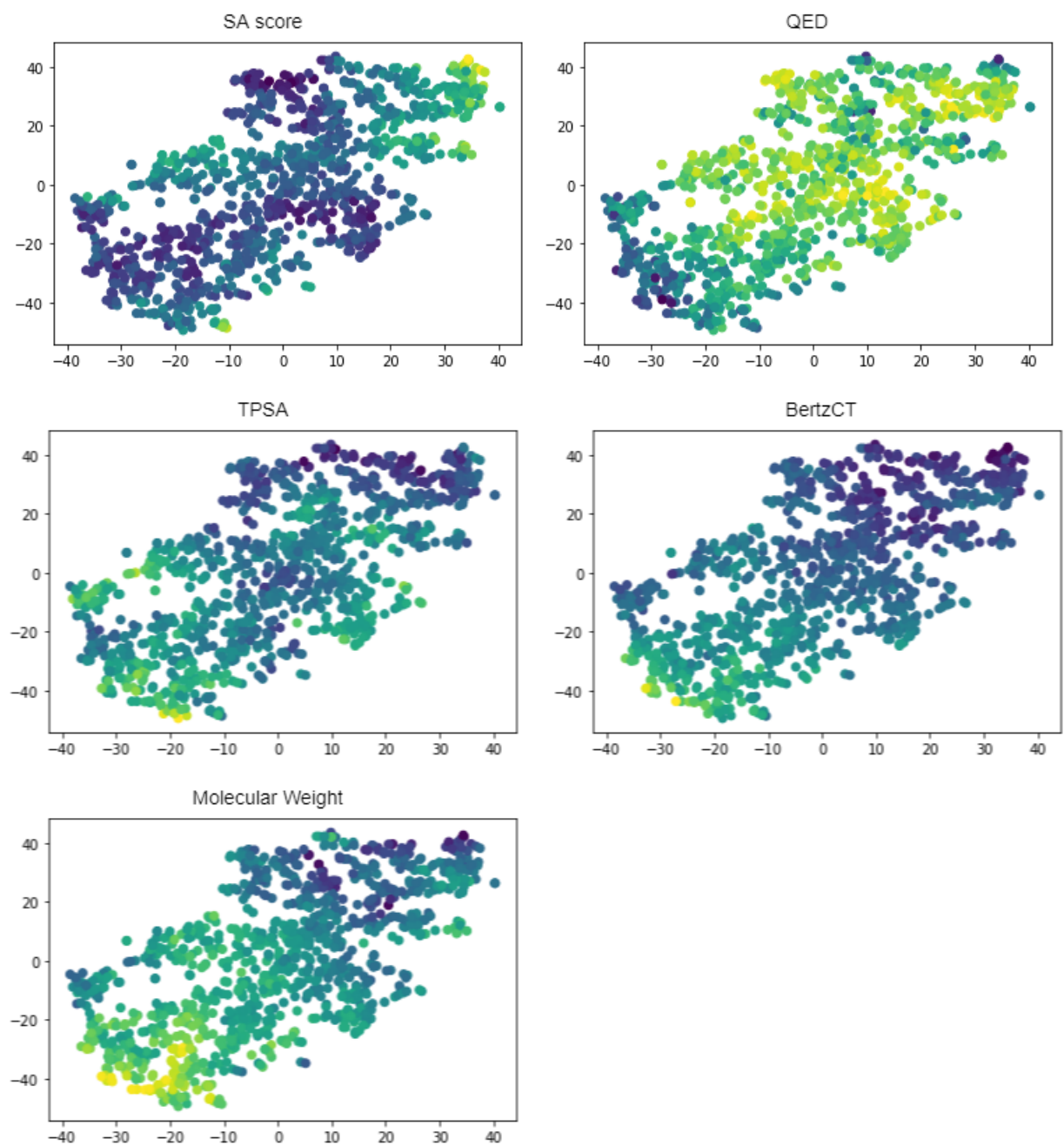

Figure 3. T-SNE of the latent vectors for our test set molecules to 2 dimensions, with scalar values of the ground truth properties as the color scale